\documentclass{aa} 

\usepackage{graphicx}

\usepackage{txfonts}

\newcommand{\teff}{$T_{\rm eff}$}
\newcommand{\logg}{$\log g$}
\newcommand{\lessim}{$^{<}_{\sim}$}

\newcommand{\pdot}{$\dot{\rm P}$}

\newcommand{\co}{$^{12}{\rm C}(\alpha,\gamma)^{16}{\rm O}$\ }

\begin{document}

\title{Two new ZZ Ceti pulsators from the HS and HE surveys
\thanks{
Based on observations collected at the Loiano Observatory, operated by the 
{\it Istituto Nazionale di Astrofisica} (INAF), and at the 
{\it Centro Astronomico Hispano Aleman} (CAHA) at Calar Alto, operated jointly
by the {\it Max-Planck Institut f\"ur Astronomie} and the {\it Instituto de 
Astrofisica de Andalucia} (CSIC).
Partially based on data obtained at the Paranal Observatory of the European 
Southern Observatory for programmes 165.H-0588 and 167.D-0407}}


\author{R. Silvotti\inst{1}, B. Voss\inst{2}, I. Bruni\inst{3}, 
D. Koester\inst{2}, D. Reimers\inst{4}, R. Napiwotzki\inst{5}
\and
D. Homeier\inst{6} 
%
%
}

\offprints{R. Silvotti}

\institute{INAF--Osservatorio Astronomico di Capodimonte, via Moiariello 16,
I-80131 Napoli, Italy\\
\email{silvotti@na.astro.it}
\and
Institut f\"ur Theoretische Physik und Astrophysik der Universit\"at Kiel,
Leibnizstra\ss e 15, D-24098 Kiel, Germany\\
\email{voss@astrophysik.uni-kiel.de,koester@astrophysik.uni-kiel.de}
\and
INAF--Osservatorio Astronomico di Bologna, via Ranzani 1,
I-40127 Bologna, Italy\\
\email{ivan.bruni@bo.astro.it}
\and
Hamburger Sternwarte, Gojenbergsweg 112, D-21029 Hamburg, Germany\\
\email{dreimers@hs.uni-hamburg.de}
\and
Centre for Astrophysics Research, University of Hertfordshire, College Lane,
Hatfield AL10 9AB, UK\\
\email{r.napiwotzki@star.herts.ac.uk}
\and
Institut f{\"u}r Astrophysik, Georg-August-Universit{\"a}t,
Friedrich-Hund-Platz 1, D-37077 G{\"o}ttingen, Germany\\
\email{derek@astro.physik.uni-goettingen.de}
%
%
}

\authorrunning{Silvotti, Voss, Bruni et al.}
\titlerunning{Two new ZZ Ceti pulsators from the HS and HE surveys}

\date{Received May, 2005; accepted ....., 2005}

\abstract{
We report the detection of nonradial $g$-mode oscillations in the DA white 
dwarfs HS~1039+4112 (B=15.9) and HE~1429$-$0343 (B=15.8) from time-series
photometry made at the Loiano 1.5~m telescope.  
The two stars were previously selected as probable pulsators based on
two-color photometry and spectral analysis respectively.
Following our temperature and surface gravity determinations, HS~1039+4112 
(\teff=11200~K, \logg=8.2) is located near the red edge of the 
ZZ Ceti instability strip, whereas HE~1429$-$0343 (\teff=11400~K, 
\logg=7.8) falls in the middle of the strip.  
Both stars show a multi-mode behavior with the main periods at about 850 
and 970~s respectively, and relatively large amplitudes ($\sim$7\% and 
$\sim$2.5\%).
}

\maketitle


\section{Introduction}

The ZZ Ceti stars (or DAVs=DA Variables) are old DA (H-rich) white dwarfs (WDs)
showing multi-periodic luminosity variations with typical periods between 
$\sim$100 and $\sim$1200 s and amplitudes between a few percent down to less
than 0.1\%, generally interpreted as nonradial $g$-mode oscillations.

The importance of studying the ZZ Ceti variables is related to the 
possibility, given by the seismological analysis, to study in detail
their interiors and to derive numerous basic stellar parameters
such as the mass, the rotation and the thickness of the external layers 
of H and He.
Due to the high stability of the stellar structure, the pulsation periods 
of the DAV white dwarfs are extremely stable in time, with values of 
\pdot~ that can be as small as $\approx$10$^{-15}$ (Mukadam et al. 
\cite{mukadam03} and references therein).
More in general, WD seismology can help to improve our knowledge of the 
poorly known rate of the \co nuclear reaction (see e.g. 
Metcalfe \cite{metcalfe03} and Straniero et al. \cite{straniero03}
and references therein), and of the WD cooling rates, including 
crystallization processes (Kanaan et al. \cite{kanaan05} and ref. therein; see
also Fontaine et al. \cite{fontaine01} for a detailed review on this argument).

The effective temperatures of the ZZ Ceti stars are confined between about 
10700 and 12100~K for a canonical surface gravity \logg$\sim$8.0 
(Bergeron et al. \cite{bergeron95}, \cite{bergeron04}; Koester \& Holberg 
\cite{koester01}; Mukadam et al. \cite{mukadam04a}).
The blue edge of the DAV instability strip can be interpreted as the limit
where the base of the surface convective zone becomes sufficiently deep 
for the local thermal time-scale to be comparable to the shortest observable 
$g$-mode pulsation period ($\approx$100 s for typical WD masses,
see Tassoul et al. \cite{tassoul90} and references therein).

This scenario is in agreement with an observed trend on temperature:
the hottest DAVs have shorter pulsation periods and lower amplitudes with
respect to the cooler ones.
More uncertain is the interpretation of the red edge: convection--pulsation 
interaction (Winget \& Fontaine \cite{winget82a}), convective mixing between 
hydrogen and helium (Bergeron et al. \cite{bergeron90}) or something 
else\hspace{0.5mm}?
This is related to the driving of the pulsations which could be based on 
the $\kappa$-$\gamma$ mechanism, active in a hydrogen layer near the 
surface (Dziembowski \& Koester \cite{dziembowski81}, Dolez \& Vauclair 
\cite{dolez81}, Winget et al. \cite{winget82b}), although an alternative 
``convective driving'' has been proposed by Brickhill (\cite{brickhill91}, 
see also Gautschy et al. \cite{gautschy96} and Goldreich \& Wu 
\cite{goldreich99}).

Recently, thanks to the results of the Sloan Digital Sky Survey, the number
of known ZZ Ceti pulsators has increased by a factor of two, reaching a total
number of 82 (Bergeron et al. \cite{bergeron04}, Mukadam et al.
\cite{mukadam04a}, Mullally et al. \cite{mullally05}).
Moreover another 5 (possibly 6) DAVs have been observed in cataclysmic 
variable systems (Woudt \& Warner \cite{woudt04}, Warner \& Woudt 
\cite{warner05}, Araujo-Betancor et al. \cite{araujo05}).
Following Mukadam et al. (\cite{mukadam04b}), it is still not clear whether 
there might be some non-variable stars within the strip, although these stars 
could simply have very low pulsation amplitudes, below the detection limit.

In this article we present two new ZZ Ceti stars, selected from the HS 
(Hamburg Schmidt, Hagen et al. \cite{hagen95}) and HE (Hamburg ESO, Christlieb
et al. \cite{christlieb01}) surveys, which bring the total number of known 
non-interacting ZZ Ceti to 84.


\section{Selection of the ZZ Ceti candidates}

The candidates were selected based on two-color photometry and spectral 
analysis.

\subsection{Two-color Str\"omgren photometry}

A list of 3000 possible cool white dwarf stars was derived from the
low-resolution photographic objective prism spectra of the Hamburg
Quasar Survey (HQS) by Homeier \& Koester (\cite{homeier01}). 
One thousand of these objects are cool (\teff\lessim16000~K) and, because of 
the large errors on the temperature (up to $\sim$6000~K), all of them are 
potential ZZ Ceti pulsators.
In order to select the best ZZ Ceti candidates, preparatory observations are 
currently being conducted, aimed to derive more precise temperatures and 
gravities for the $\approx$~500 brightest stars (B$<$16.5~mag).

We obtained photometry in the Str\"omgren $u$, $b$, $y$ and Johnson $I$ bands 
at the Calar Alto 2.2m telescope using the BUSCA four-channel CCD camera
(Reif et al. \cite{reif99})~\footnote{See also the web pages of BUSCA at 
{\it http://www.caha.es/guijarro/BUSCA/intro.html}}.
The colours $u-b$ and $b-y$ show a strong dependency on \teff\ and \logg, 
which is strongest in the region of the ZZ Ceti instability strip.
In the course of our analysis, we noticed technical problems with the BUSCA 
$y$ filter as well as an even stronger temperature dependency of $b-I$ than 
that of $b-y$. 
Therefore we use the $u-b$ vs. $b-I$ two-color diagram for the analysis 
of our data.

To derive atmospheric parameters from the two-color data, we compare them to a
grid of synthetic colors. 
This grid was computed from Koester DA model atmospheres (described e.g. in 
Finley et al. 1997) by convolving the model spectra with the BUSCA filter 
transmission curves. 
Temperatures and gravities were determined by interpolating linearly between 
the grid points. 
The observational and calibration errors of the colors are $\sim0.02$~mag, which
translates into an error of $\sim\pm0.1$ in \logg\ and $\sim\pm300$~K in \teff.
This temperature error is small compared to the width of the instability strip,
and thus most objects are clearly placed either outside or inside the 
instability strip by the photometry. 

\begin{figure}[|b,t]
\label{fig:grid}
\vspace{65mm}
\includegraphics{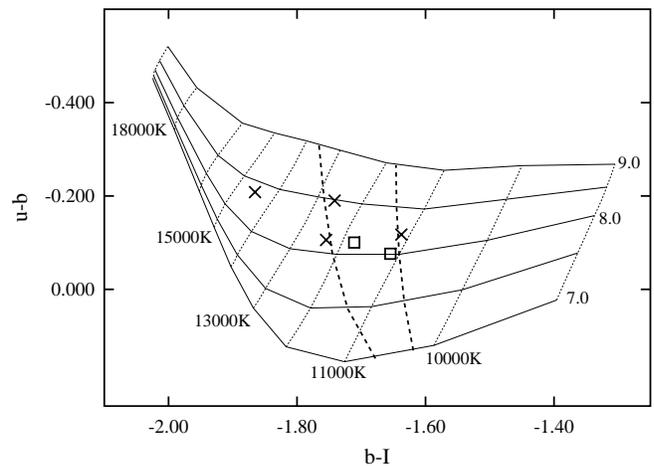}
\caption{Two-color data and the grid of synthetic colors from model
 atmospheres. Solid lines are lines of constant gravity and dotted lines are 
 lines of constant temperature (\teff=20000~K, 18000, 16000, 15000
 and down to 9000 in steps of 1000~K). 
 Dashed lines show the boundaries of the instability strip,
 according to Bergeron et al. (2004). 
 The crosses represent the four stars selected as possible pulsators:
 from left to right HS~1610+1639, HS~1443+2934, HS~1253+1033 and 
 HS~1039+4112.
 The squares are two previously known ZZ Ceti stars, GD~244 (left) and
 PG~2303+242, that were included in the preparatory observations as a check 
 of our methods.}
 \end{figure}

Four objects that were selected as potential pulsators in this way are
studied here. 
Their two-color data are plotted in Fig.~1, together with the 
temperature/gravity grid and the instability strip edges, taken from Bergeron
et al. (2004). 
For one of the new pulsators, HS~1039+4112 (RA$_{2000}$=10 42 33.5, 
DEC$_{2000}$=+40 57 16, B=15.9~mag), we find \teff=11200$\pm$270~K and 
\logg=8.2$\pm$0.12.
The parameters of the three remaining objects are given in Table~2.

The selection of the object HS~1610+1639 is problematic:
using the now obsolete $b-y$ data, we obtained \teff=11700~K and \logg=8.2,
corresponding to the middle of the instability strip.
This is the reason why HS~1610+1639 was included in the sample observed
at Loiano.
However, after the time series observations at Loiano had been made,
the re-analysis of the data, now using the $b-I$ color, has yielded a higher 
temperature for this object, \teff=14500~K (and \logg=8.4), 
definitively too hot for ZZ Ceti pulsations.

\subsection{Spectral analysis}

The other new pulsator, HE~1429$-$0343 (RA$_{2000}$=14 32 03.2, 
DEC$_{2000}$=--03 56 38, B=15.8), was selected based on high-resolution
spectra from the SPY survey (Napiwotzki et al. 2003).
An analysis of these spectra for 754 white dwarfs, many of which are first 
spectroscopic confirmations of WDs from the HS and HE surveys, was conducted 
by Voss et al. (2005, in preparation). 
Details about theoretical spectra and fitting procedures can be found in 
Finley et al. (1997).
Fig.~2 displays the model atmosphere fit for HE~1429$-$0343, from which we 
obtain \teff=$11434\pm 36$ K and \logg=$7.82\pm 0.02$.  

\begin{figure}[|h]
\label{fig:spectrum1}
\vspace{64mm}
\includegraphics{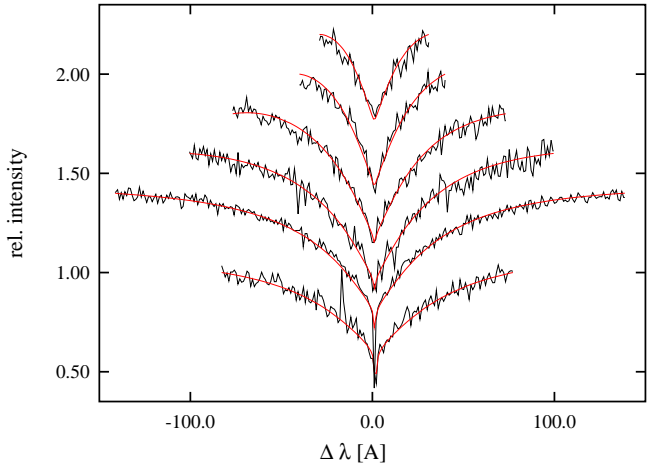}
\caption{The SPY spectrum and fit of HE~1429$-$0343. 
 The spectrum was rebinned to a resolution of 1\AA~ for this plot, the fit was
 done with the original resolution of 0.1\AA.
 The deep core of H$\beta$ is not real but is caused by a cosmic, which 
 however does not significantly affect the fitting process.}
\end{figure}

\begin{figure}[|h,t,b]
\label{fig:spectrum2}
\vspace{64mm}
\includegraphics{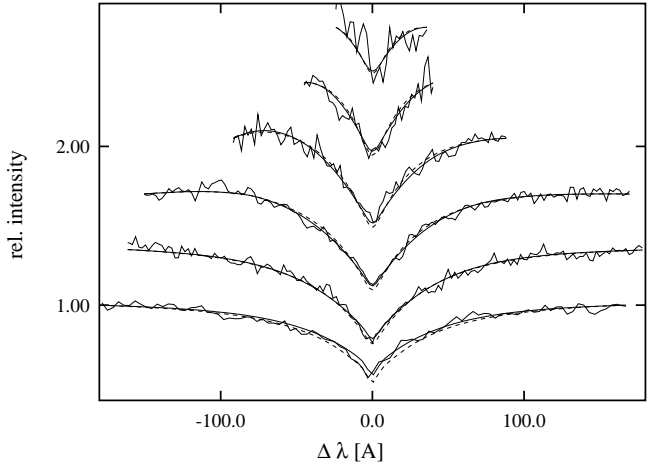}
\caption{Spectrum of HS\,1641+7132 and fits for \teff\,=\,15\,350\,K, 
 \logg\,=\,7.9 (solid line), and for 11\,800\,K, \logg\,=\,8.1 (dashed)}
 \end{figure}

The other object selected from spectroscopy is HS~1641+7132, one of the 12 
possible DAVs identified from medium-resolution spectroscopy by Homeier et al.
(\cite{homeier98}).
HS\,1641+7132 was selected for time-series photometry because of its 
atmospheric parameters, \teff=$11600\pm 80$~K and \logg=$8.0\pm 0.06$, 
which correspond to the middle of the instability strip.
While the formal errors are small despite the low $S/N$ and the resolution
(6\,{\AA}), line profile fits near the Balmer line maximum are known to 
easily lead to degenerate solutions for \teff~ and \logg. 
Indeed, upon re-analyzing the original data from \cite{homeier98}, we found 
that even small changes to the startup values of the fitting procedure would 
produce temperatures as much as 200\,K apart, with a mean value of 11800\,K.
In addition, more important, we found a second solution for the fit at 
much higher temperature \teff=$15350\pm 300$ K and \logg=$7.88\pm 0.06$, 
with a reduced  $\chi2$\,=\,1.26 as compared to 1.45 for the cooler solution.
The two fits of the spectrum are shown in Fig.~3.

\section{Time-series photometry}
 
The time-series observations were performed in June 2004 at the 1.5~m Loiano 
telescope using the BFOSC~\footnote{Bologna Faint Object Spectrograph \& Camera,
see the web site of the Loiano Observatory at 
{\it http://www.bo.astro.it/loiano/} for more details.}
instrument without any filter.
A recent version of the software allows to launch a series of exposures
where the CCD is read only in a small box around the target, in order
to decrease the read-out time.
With a box of typically $\sim$300$\times$400 pixels, we were able
to measure at least two reference stars in each field.
The integration times used were between 12 and 18~s, giving an effective
resolution of about 24 to 30~s, because of the CCD reading plus some extra
time due to internal controls of the software.

The data were then reduced using IRAF~\footnote{Image Reduction and Analysis 
Facility, written and supported by the IRAF programming group at the National 
Optical Astronomy Observatories (NOAO) in Tucson, Arizona 
({\it http://iraf.noao.edu/}).} aperture photometry.
The flux of each target was divided by a combination of the reference stars
with different weights, and then normalized to the mean ratio.
Finally a residual extinction correction was applied and the times were
converted to Barycentric Julian Date (BJD), using the algorithm of
Stumpff (\cite{stumpff80}).

The light curves of the two new ZZ Ceti stars are shown in Fig.~4 and 5.
The variable amplitudes in the light curves, presumably due to beating effects,
suggest that both targets have more than one excited pulsation mode.
Each target was observed in two different nights to confirm the pulsations.
In this way, joining together the two runs, the frequency resolution of
the Fourier Transform is improved and it is possible to make a first
separation of the different peaks (Fig.~4 and 5).
The light curve of HS~1039+4112 has the typical large amplitude of many 
of the cool ZZ Ceti; moreover the spectrum shows the 
possible presence of the first and second harmonic of the main peak 
at 1.181 mHz (846.6~s).
In the case of HE~1429$-$0343, which is located more close to the center 
of the instability strip, the pulsation amplitude is lower.
Close to the highest peak of the spectrum at 1.032 mHz (969.0~s)~\footnote{The
amplitude of this mode, however, is not the highest following the results of 
the sinusoidal fits, as reported in Table~1.}
there are at least two other signals at 0.922 and 1.206 mHz (1084.9 and 829.3~s
respectively).

\begin{figure}[h]
\label{fig:hs1039}
\vspace{150mm}
\includegraphics{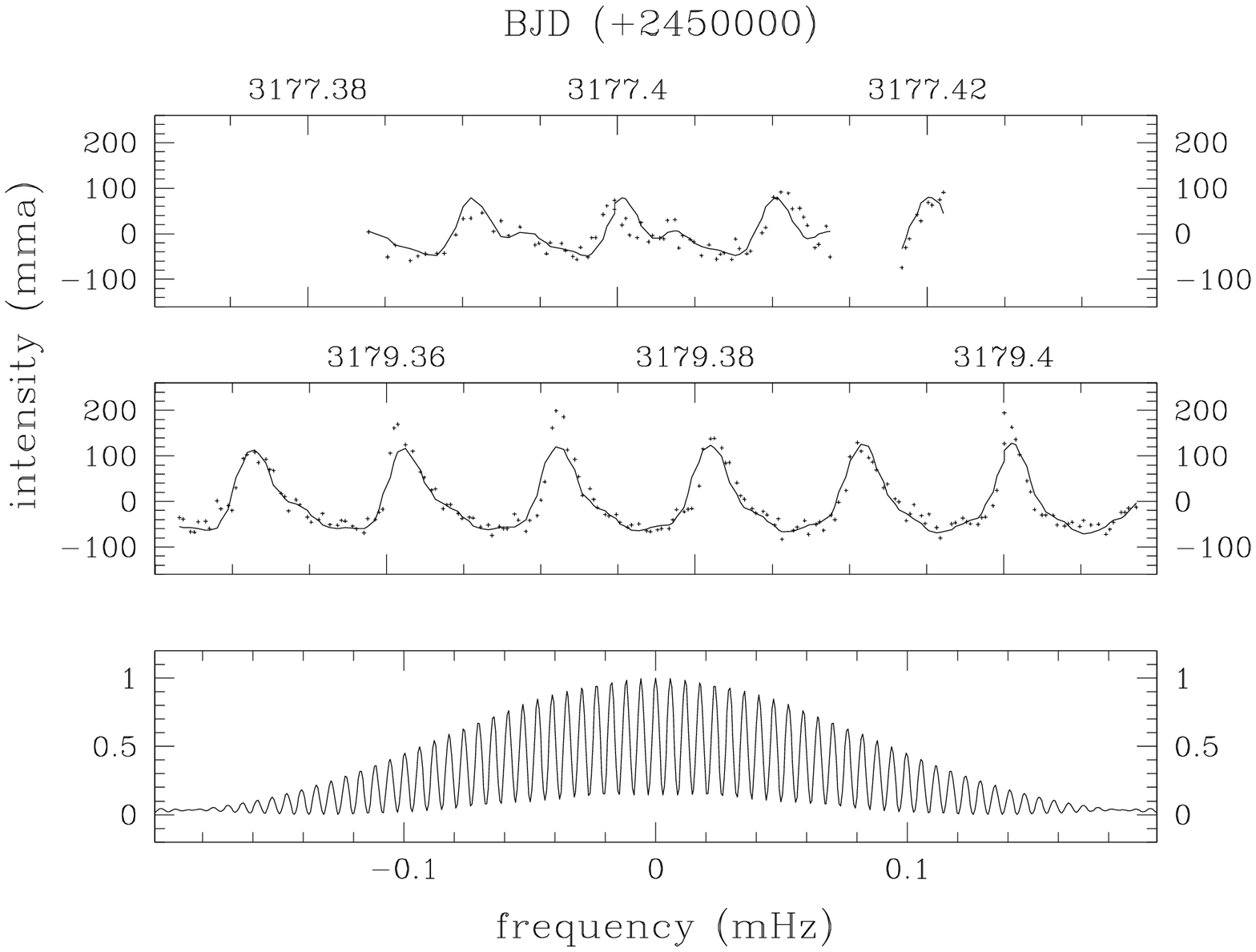}
\includegraphics{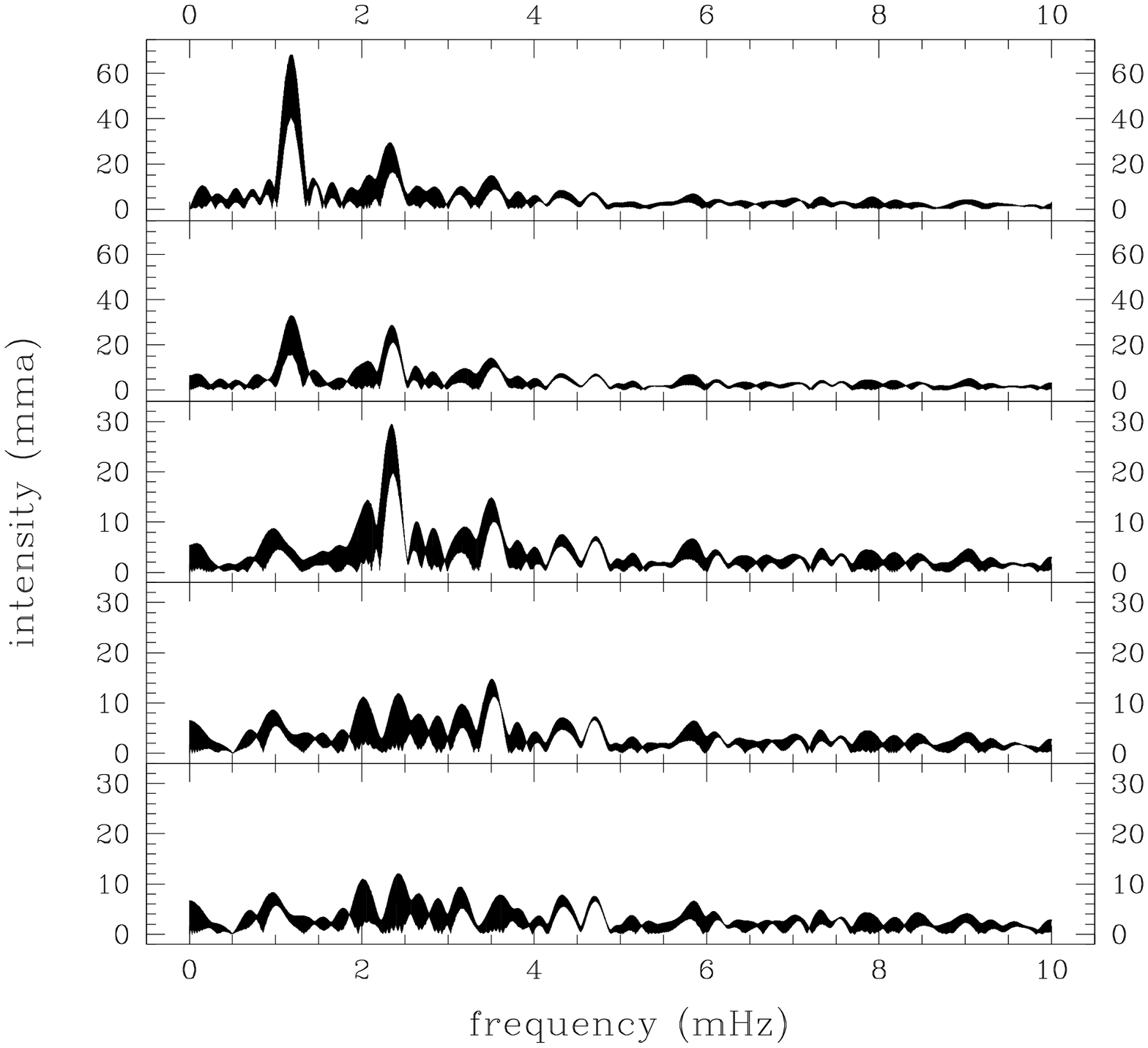}
\caption{Light curves, spectral window and amplitude spectrum of HS~1039+4112
 (1~mma = 0.1\% change in intensity).
 The observations were performed in June 20 and 22, 2004.
 The amplitude spectrum is obtained joining together the two runs in order
 to increase the frequency resolution.
 The lower panels show the behavior of the spectrum when we subtract, at each
 iteration, the highest peak (pre-whitening); note that the intensity scale
 is not the same in all panels.}
\end{figure}

\begin{table}[|h,b] 
\label{tbl:fit}
\begin{center}
\caption{Results of the non-linear least-squares fits} 
\begin{tabular}{lcccccccccccc} 
\hline
\bf Name & & \bf F(mHz) & \bf P(s) & \bf A(mma) \\
\hline
\object{HS~1039+4112}   & F1  &  1.169  &  855.5  &  55.2 \\ 
                        &     & [1.181] & [846.6] & [71.8]\\ 
                        & F2  &  1.194  &  837.3  &  26.0 \\ 
                        &     &         &         & [30.7]\\ 
			& 2F1 &  2.338  &  427.8  &  29.1 \\ 
			& 3F1 &  3.507  &  285.1  &  14.6 \\   
\hline
\object{HE~1429$-$0343} & F1 & 0.922 & 1084.9 & 16.3 \\ 
                        & F2 & 1.032 &  969.0 & 12.7 \\  
			& F3 & 1.206 &  829.3 & 18.3 \\ 
			& F4 & 2.222 &  450.1 & 10.2 \\  
\hline
\end{tabular}
\end{center}
\end{table}

\begin{figure}[h]
\label{fig:he1429}
\vspace{150mm}
\includegraphics{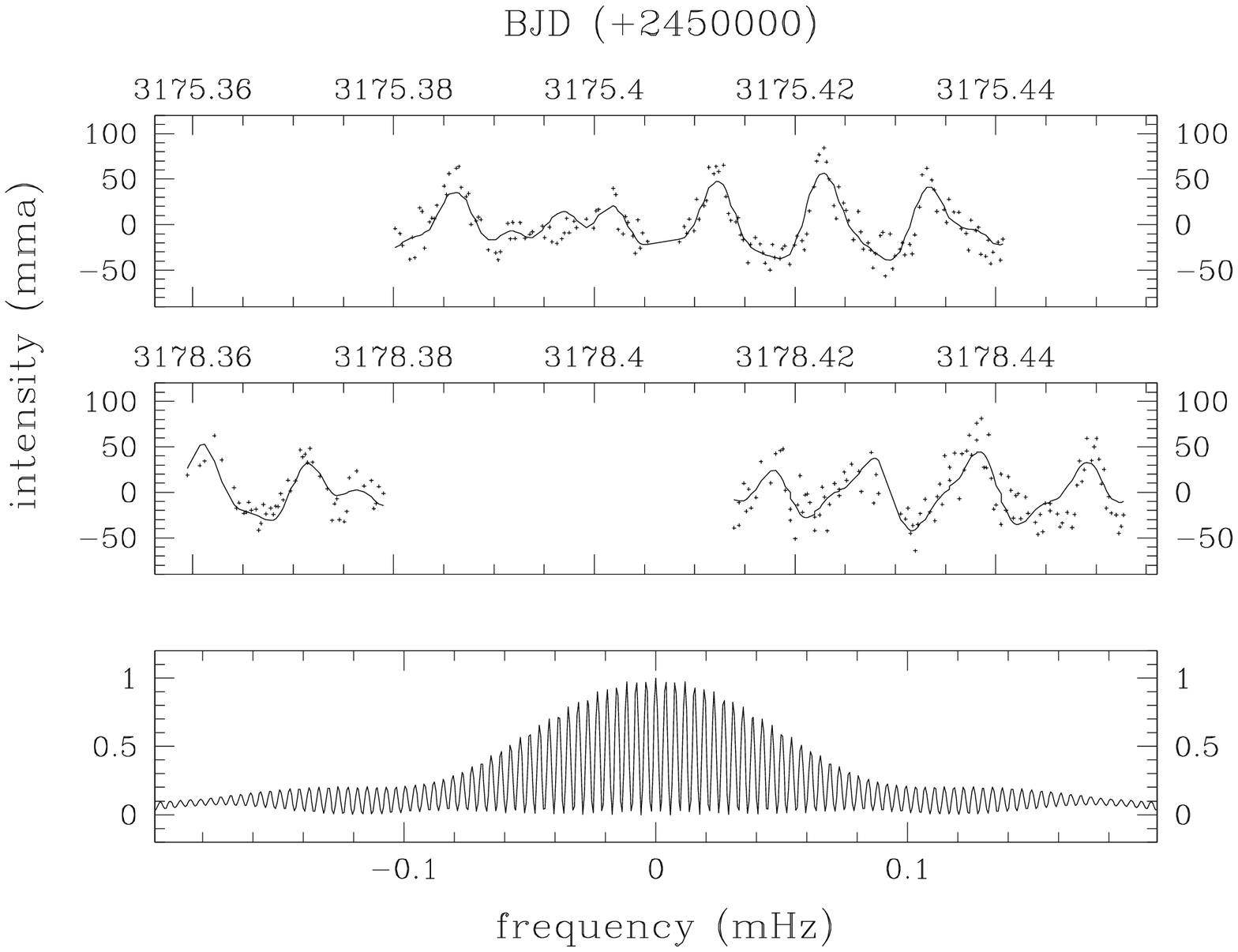}
\includegraphics{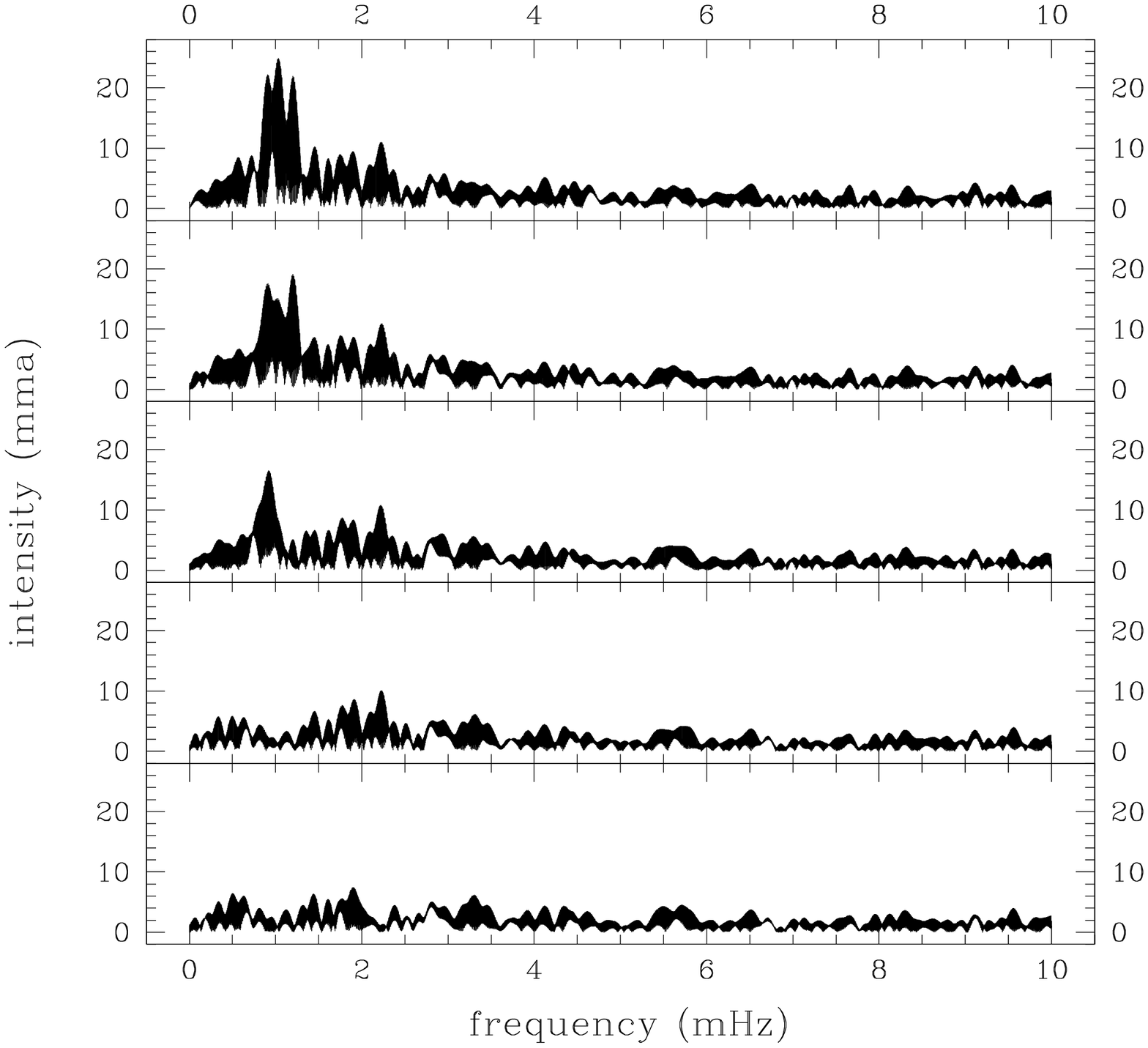}
\caption{Same as Fig.~4, but for HE~1429$-$0343.
 The observations of HE~1429$-$0343 were performed in June 18 and 21, 2004.}
\end{figure}

In Table~1
we report an attempt to fit the observed light curves (both nights together) 
with sinusoidal fits; the results of the fits are also shown in the upper 
panels of Fig.~4 and 5.
However, due to the poor coverage, the pulsation spectra are not resolved  
and therefore these results should be taken with some prudence, in particular 
regarding the frequencies with lower amplitudes (3F1 and F4), which are quite 
close to the noise level.
Concerning HS~1039+4112, the first and second harmonics suggest 
that the true value of the main frequency could be 1.169 mHz, 
whereas the value of 1.181 detected in the power spectrum could correspond 
to the 1-day alias (1.169 + 0.012 mHz).
Indeed, using 1.169 mHz, the least-squares fit is more stable and therefore
we adopt this solution in Table~1 (the alternative solution is also given
in square brackets for completeness).
In the case of HE~1429$-$0343, the results of the sinusoidal fits favor
a higher amplitude of F3 and F1 respect to F2 which, however, is the highest 
peak in the Fourier spectrum (24.9 mma).

\section{The ZZ Ceti instability strip updated}

In Fig.~6
the location of the two new ZZ Ceti stars in a (\teff, \logg) plane
is shown and compared with the edges of the observational ZZ Ceti instability 
strip as defined by Bergeron et al. (\cite{bergeron04}).
Both stars are within the strip, although HS~1039+4112 is very close to the 
red edge.

\begin{figure}[h]
\label{fig:davs}
\vspace{84mm}
\includegraphics{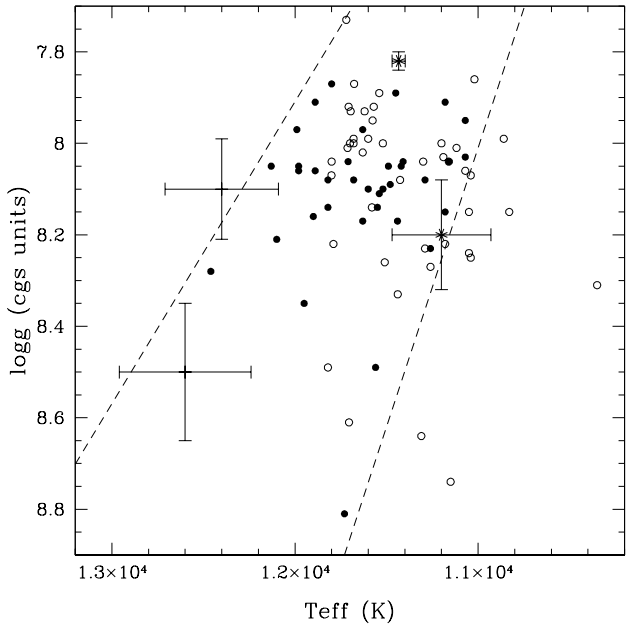}
\caption{The observational ZZ Ceti instability strip updated: the two new 
pulsators are represented by the two asterisks with error bars, whereas the two
objects close to the blue edge are HS~1253+1033 and HS~1443+2934 (from left 
to right), for which no pulsations were found.
The filled circles are the known ZZ Ceti stars from Bergeron et al. (2004),
the open circles are the new pulsators discovered from the SDSS 
(Mukadam et al. 2004a; Mullally et al. 2005).}
\end{figure}

\begin{table*}[t] 
\label{tbl:nonpulsators}
\begin{center}
\caption{Objects which do not show pulsations} 
\begin{tabular}{lcccccccc} 
\hline
\bf Name & \bf RA$_{2000}$ & \bf DEC$_{2000}$ & \bf B   & \bf \teff
         & \bf \logg       & \bf UT date      & \bf D   & \bf L \\
         &                 &                  &         & (K)
	 & (cgs)           & at start         & (hours) & (mma) \\
\hline
\object{HS~1253+1033} & 12 56 28.5 & +10 17 09 & 14.4 & 12600
                      & 8.5        & 16/06/04  & 1.6  & 5.3 \\
\object{HS~1443+2934} & 14 45 28.5 & +29 21 31 & 14.5 & 12400
                      & 8.1        & 22/06/04  & 2.0  & 1.5 \\
		      &  	   &           &      &
		      &  	   & 23/06/04  & 1.5  & 1.2 \\
\object{HS~1610+1639} & 16 13 02.4 & +16 31 56 & 15.8 & 14500 
                      & 8.4        & 17/06/04  & 3.4  & 2.0 \\   
\object{HS~1641+7132} & 16 41 00.0 & +71 26 59 & 16.7  
                      & \hspace{1.5mm}15300$^{\ast}$       
                      & \hspace{1.5mm}7.9$^{\ast}$ 
		      & 23/06/04   & 1.4       & 4.6 \\ 
\hline
\end{tabular}
\end{center}
$^{\ast}$An alternative solution for this star gives 
\teff=11600 K and \logg=8.0 (see section 2.2 for more details).
\end{table*}

On the other hand, if we consider the other four targets observed in the same 
run at Loiano, none of them shows pulsations. 
Their upper limits (i.e. the amplitude of the highest peak in the region of 
interest of their temporal spectra) are reported in Table~2,
together with their B magnitudes,
effective temperatures and surface gravities, 
date and duration of the observations.

Concerning HS~1641+7132, its non-variability is compatible with the 
higher temperature solution which was found upon re-analyzing the spectroscopic
data, as described in section 2.2.
Therefore we conclude that this star is likely to be a non-variable DA 
with \teff~ between 15000 and 16000\,K. 
The same applies to HS~1610+1639, which is clearly far from the blue edge of 
the instability strip, having an effective temperature of about 14500\,K.

Finally, for what concerns the other two stars which do not show luminosity 
variations, HS~1253+1033 and HS~1443+2934, both lie near the blue edge.
The case of HS~1253+1033 is the most intriguing as this star apparently falls
inside the observational strip of Bergeron et al. (\cite{bergeron04}).
However, even in this case, the relatively large uncertainties in \teff~ and
\logg~ can not completely exclude the possibility that HS~1253+1033 is actually
slightly beyond the blue edge.
%
%
Moreover, looking at Fig.~6, both stars would be outside the observational 
strip as defined by the SDSS sample alone. Considering that we used the same 
Koester model atmospheres as used by the SDSS team, the stability against
$g$-modes of HS~1253+1033 and HS~1443+2934 appears still less surprising.

\begin{acknowledgements}

R.S. acknowledges support from the funds COFIN ``Astrosismologia''
(PI L.~Patern\`o).
D.K. and B.V. thank the Deutsche Forschungsgemeinschaft (DFG) for their
support (KO738/21-1, KO738/22-1, KO738/23-1).
R.N. acknowledges support by a PPARC Advanced Fellowship.
D.H. thanks the DFG for grant KO738/10.
The authors wish to thank Anjum Mukadam for a very accurate referee report 
that has contributed to improve the quality of this article.

\vspace{2mm}
\noindent
{\it R.S. ringrazia il padre Giovanni nel giorno del suo compleanno, 
a met\`a del run osservativo, e ricorda il felice festeggiamento insieme 
il 20 Giugno 2004}.

\end{acknowledgements}

\end{document}